\documentclass[a4paper,11pt]{article}
\usepackage{pos}

\usepackage{amsfonts} 
\usepackage{amsmath} 
\usepackage{subcaption}
\usepackage{array} 
\usepackage{bm} 
\usepackage{latexsym} 
\usepackage{longtable} 
\usepackage{hyperref} 
\usepackage{multirow}
\usepackage{xcolor}
\usepackage{bbold}
\usepackage{lipsum}
\usepackage{mathtools}
\usepackage{comment}
\usepackage{paralist}
\usepackage{placeins}
\usepackage{verbatim}
\usepackage{adjustbox}
\usepackage{enumitem}
\usepackage{caption}

\graphicspath{{./figs/}}

\allowdisplaybreaks

\definecolor{HLBlue}{HTML}{6599FF}
\definecolor{HLOrange}{HTML}{FF6600}

\newcommand{\BtoDst}{\bar{B} \to D^\ast \ell \bar{\nu}}

\title{Current progress on the semileptonic form factors for $\BtoDst$ decay using the Oktay-Kronfeld action}
\ShortTitle{Data analysis on 2pt and 3pt function}

\author[a,1]{Tanmoy Bhattacharya}
\author*[b,1]{Benjamin J.~Choi}
\author[a,1]{Rajan Gupta}
\author[c,1]{Yong-Chull Jang}
\author[b,1]{Seungyeob Jwa}
\author[b,1]{Sunghee Kim}
\author[b,1]{Sunkyu Lee}
\author[b,1]{Weonjong Lee}
\author[d,1]{Jaehoon Leem}
\author[b,1]{Jeonghwan Pak}
\author[e,1]{Sungwoo Park}

\affiliation[a]{Theoretical Division T-2, Los Alamos National
  Laboratory, Los Alamos, NM 87545, USA}

\affiliation[b]{Lattice Gauge Theory Research Center, CTP and FPRD,
  Department of Physics and Astronomy,\\ Seoul National University,
  Seoul 08826, South Korea}

\affiliation[c]{Physics  Department, Brookhaven National Lab, Upton,
      NY 11973, USA}

\affiliation[d]{Computational Science and Engineering Team, Innovation
  Center, Samsung Electronics, Hwaseong, Gyeonggi-do 18448, South
  Korea}

\affiliation[e]{Lawrence Livermore National Lab, 7000 East Ave,
  Livermore, CA 94550, USA}

\note{The LANL-SWME Collaboration}

\emailAdd{benjaminchoi@snu.ac.kr}

\emailAdd{wlee@snu.ac.kr}

\abstract{We present recent progress in calculating the semileptonic
  form factors $h_{A_1}(w)$ for the $\BtoDst$ decays.
  We use the Oktay-Kronfeld (OK) action for the charm and bottom
  valence quarks and the HISQ action for light quarks.
  We adopt the Newton method combined with the scanning method to
  find a good initial guess for the $\chi^2$ minimizer in the fitting
  of the 2pt correlation functions.
  The main advantage is that the Newton method lets us to consume all
  the time slices allowed by the physical positivity.
  We report the first, reliable, but preliminary results for
  $h_{A_1}(w)/\rho_{A_1}$ at zero recoil ($w=1$).
  Here we use a MILC HISQ ensemble ($a = 0.12$ fm, $M_{\pi}$ = 220
  MeV, and $N_f = 2 + 1 + 1$ flavors). }

\FullConference{The 40th International Symposium on Lattice Field Theory (Lattice 2023)\\
July 31st - August 4th, 2023\\
Fermi National Accelerator Laboratory\\}

\begin{document}
\maketitle

%
%
\section{Introduction}
\label{sec:intr}
We present update of data analysis on the 2pt and 3pt
correlation functions to obtain the semileptonic form factors for the
$\BtoDst$ decays.
We adopt the Newton method combined with the scanning method
\cite{Bhattacharya:2021peq} to find a good initial guess for the
$\chi^2$ minimizer in the fitting of the 2pt correlation functions.
We find that the multiple time slice combinations help to distinguish
the global minimum of $\chi^2$ and its local minima reliably.
The Newton method leads to a self-consistent fit which consumes all
the time slices allowed by the physical positivity \cite{ Luscher:1976ms, Luscher:1984is}.
The results of data analysis on the 2pt correlation functions are
used as inputs to the fitting of the 3pt correlation functions.
As a result, we report the first, reliable, but preliminary results
for $h_{A_1}(w)/\rho_{A_1}$ at zero recoil ($w=1$), obtained using the
MILC HISQ ensemble in Table \ref{tab:ensembles}.
\begin{table}[!h]
  \renewcommand{\arraystretch}{1.2}
  \center
  \resizebox{0.9\textwidth}{!}{
    \begin{tabular}{ @{\quad} c @{\quad}|@{\quad} c @{\quad}|@{\quad} c @{\quad}|@{\quad} c @{\quad}|@{\quad} c @{\quad} c @{\quad} c @{\quad}}
      \hline\hline $a$ (fm) & $N_f$
      & $N_s^3 \times N_t$
      & $M_\pi$ (MeV)
      & $am_l$ & $am_s$ & $am_c$ \\
    \hline
    0.1184(10) & $2+1+1$ & $32^3\times 64$ & 216.9(2) & 0.00507 & 0.0507 & 0.628
    \\
    \hline\hline
    \end{tabular}
  } 
  \caption{ Details on the MILC HISQ ensemble used for the
    numerical study \cite{Bazavov:2012xda}.}
  \label{tab:ensembles}
\end{table}

%
%

%
%
\section{Flow chart of the data analysis}
\label{sec:flow}
%
%
%
The 2-point (2pt) correlation function is defined as
\cite{Bazavov:2011aa},
\begin{align}
  C(t) &= \sum_{\alpha = 1}^{4} \sum_{\bf{x}} \left\langle
  \mathcal{O}^{\dagger}_\alpha(t,{\bf x}) \mathcal{O}_\alpha(0)
  \right\rangle \,, \qquad
  \label{eq:corr-2pt}
\end{align}
where an interpolating operator for heavy-light mesons
$\mathcal{O}_\alpha(t,{\bf x})$ is
\begin{align}
  \mathcal{O}_\alpha(t,{\bf x}) &= \left[ \bar{\psi}(t,{\bf x})
    \, \Gamma \, \Omega(t,{\bf x}) \right]_\alpha \chi(t,{\bf x}) \,, \qquad
  \Omega(t,{\bf x}) \equiv \gamma_{1}^{\;x_1} \gamma_{2}^{\;x_2}
  \gamma_{3}^{\;x_3} \gamma_{4}^{\;t} \,.
\end{align}
Here, $\Gamma = \gamma_5$ ($\Gamma = \gamma_j$) for the pseudoscalar
(vector) meson.
Here, $\psi$ is a heavy quark field in the OK action
\cite{Oktay:2008ex}, $\chi$ is a light quark field in the HISQ
staggered action \cite{Follana:2006rc}, and the subscript $\alpha$
represents taste degrees of freedom for staggered quarks.
We measure meson propagators (\emph{i.e.} 2pt correlators in
Eq.~\eqref{eq:corr-2pt}) on the lattice.
In the lattice QCD, the lattice Hilbert space consists of states of
quarks and gluons, but the physical Hilbert space consists of states
of hadrons.
Hence, in order to extract physical information on hadronic states from
the 2pt correlator, we use the spectral decomposition in the physical
Hilbert space to obtain the fitting functional form $f(t)$.
The fitting function of the $m+n$ fit is
\begin{align}
  f(t) & = g(t) + g(T-t)\,,
  \nonumber \\
  g(t) &= A_0 \, e^{-E_0 \, t} \left[ 1 + R_{2} \, e^{-\Delta E_{2} \,
      t} \left( 1 + R_{4} \, e^{-\Delta E_{4} \, t} \left( \cdots
    \left( 1 + R_{2m-2} \, e^{-\Delta E_{2m-2} \, t} \right) \cdots
    \right) \right) \right.
    \nonumber \\
    & \hphantom{=} \left. - (-1)^{t} \, R_{1}
      \, e^{-\Delta E_{1} \, t} \left( 1 + R_{3} \, e^{-\Delta E_{3}
        \, t} \left( \cdots \left( 1 + R_{2n-1} \, e^{-\Delta E_{2n-1}
        \, t} \right) \cdots \right) \right) \right]
  \label{eq:fit-func}
\end{align}
where $\Delta E_i \equiv E_i - E_{i-2}$, $E_{-1}\equiv E_0$\,, $R_i
\equiv \dfrac{ A_i }{ A_{i-2} }$ and $A_{-1} \equiv A_0$.
The subscript ${}_0$ represents the ground state in the physical Hilbert
space.
Hence, $A_0$ and $E_0$ represents the amplitude and energy of the ground
state.
In the $m+n$ fit, $m$ ($n$) is the number of even (odd) time-parity
states, which are kept in the fitting, while higher excited states in
each time-parity channel are truncated.

\begin{table}[t!]
  \renewcommand{\arraystretch}{1.2}
  \begin{subtable}{\linewidth}
    \centering
    \vspace*{-5mm}
    \begin{tabular}{@{\quad}c@{\quad} | @{\quad}l@{\;\;} | @{\quad}l@{\quad}}
      \hline
      \hline
      Symbol & Description & Example
      \\ \hline
      $\bar{\lambda} \pm \sigma(\lambda)$
      & average and error of a fit parameter $\lambda$
      &  $\bar{A}_0 \pm \sigma(A_0), \ldots$
      \\
      $\lambda_p \pm \sigma_p(\lambda)$
      & prior information on a fit parameter $\lambda$
      & $[A_0]_p \pm \sigma_p(A_0), \ldots$
      \\
      $\sigma_p^{\text{mf}}(\lambda)$
      & $\sigma_p(\lambda)$ obtained by the maximal fluctuation of data
      & $\sigma_p^{\text{mf}}(A_0), \ldots$
      \\
      $\sigma_p^{\text{sc}}(\lambda)$
      & $\sigma_p(\lambda)$ obtained by the signal cut ($\sigma_p^{\text{sc}}(\lambda) = |\bar{\lambda}_p|$)
      & $\sigma_p^{\text{sc}}(A_0), \ldots$
      \\
      $\sigma_p^{\max}(\lambda)$
      & $\min \bigl( \;\sigma_p^{\text{mf}}(\lambda), \;\sigma_p^{\text{sc}}(\lambda) \bigr)$
      & $\sigma_p^{\max}(A_0), \ldots$
      \\
      $\sigma_p^{\text{opt}}(\lambda)$
      & optimal prior width of $\lambda$ in Eq.~\eqref{eq:opt-pwidth-def}
      & $\sigma_p^{\text{opt}}(A_0), \ldots$
      \\
      $\sigma_\sigma (\lambda)$
      & error of error for $\lambda$, that is, error of $\sigma(\lambda)$
      & $\sigma(A_0) \pm \sigma_\sigma(A_0), \ldots$
      \\
      \hline
      \hline
    \end{tabular}
    \caption{Notation and convention}
    \label{subtab:symbol-1}
  \end{subtable}
  \begin{subtable}{\linewidth}\centering
    \begin{tabular}{@{\quad}c@{\quad} | @{\quad}l@{\quad}}
      \hline
      \hline
      Symbol & Description
      \\ \hline
      $\sigma(A_0; \max)$
      & $\sigma(A_0)$ when we set $\sigma_p(A_0) = \sigma_p^{\max}(A_0)$
      and $\sigma_p(E_0) = \sigma_p^{\max}(E_0)\}$
      \\
      $\sigma(E_0; \max)$
      & $\sigma(E_0)$ when we set $\sigma_p(A_0) = \sigma_p^{\max}(A_0)$
      and $\sigma_p(E_0) = \sigma_p^{\max}(E_0)\}$
      \\
      \hline
      \hline
    \end{tabular}
    \caption{Notation used for stability tests}
    \label{subtab:symbol-2}
  \end{subtable}
  \caption{Notation and convention used in this article.}
  \label{tab:symbol}
\end{table}

Our notation and convention is described in Table \ref{tab:symbol}.

To determine fit parameters, $A_0$, $E_0$, $\{R_j$, $\Delta E_j\}$, we
use sequential Bayesian method.
We obtain fit parameters which minimizes the $\chi^2$.
Using the fit function given in Eq.~\eqref{eq:fit-func}, we adopt the
Broyden-Fletcher-Goldfarb-Shanno (BFGS) algorithm \cite{
  Broyden:1970bro, Fletcher:1970fle, Goldfarb:1970gol, Shanno:1970sha}
for the $\chi^2$ minimizer, which belongs to the quasi-Newton method.
The quasi-Newton method needs an initial guess for the fit parameters.
Here, we denote the initial guess as $A_0^g$, $E_0^g$, $\{ R_j^g$,
$\Delta E_j^g \}$.
The superscript ${}^g$ represents the ``initial guess''.
A good initial guess reduces the number of iterations in the BFGS
minimizer, which saves the computing cost dramatically \cite{
  Bhattacharya:2021peq}.
In order to find a good initial guess directly from the data, we use
the multi-dimensional Newton method \cite{ Press:2007nr,
  Broyden:1965br} combined with the scanning method \cite{
  Bhattacharya:2021peq}.
We present the flow chart for the sequential Bayesian method in the
following to describe the logistics.

\begin{figure}[tb]
  \begin{subfigure}{0.49\textwidth}
    \begin{center}
      \vspace*{-5mm}
      \includegraphics[width=\linewidth]
                      {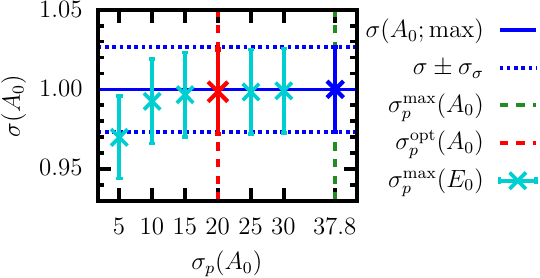}
    \end{center}
    \vspace{-1.0em}
    \caption{$\sigma_p^{\text{opt}}(A_0)$ in the unit of $\sigma(A_0;\max)$.}
    \label{fig:subfig-2+1-stab-A0-1}
  \end{subfigure}
  \hfill
  \begin{subfigure}{0.49\textwidth}
    \begin{center}
      \vspace*{-5mm}
      \includegraphics[width=\linewidth]
                      {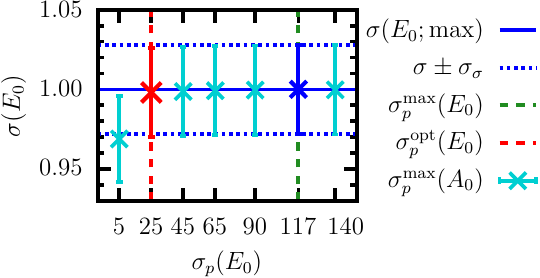}
    \end{center}
    \vspace{-1.0em}
    \caption{$\sigma_p^{\text{opt}}(E_0)$ in the unit of $\sigma(E_0;\max)$.}
    \label{fig:subfig-2+1-stab-E0-1}
  \end{subfigure}
  \caption{ Results of the stability tests to find optimal prior
    widths for the 2+1 fit.}
  \label{fig:stab-2+1-1}
\end{figure}

%
%
\begin{enumerate}[ label=\textbf{Step-\arabic*}, ref=\textbf{Step-\arabic*},
    itemsep=0.5pt]
\item \label{it:sb-1} Do the 1st fit. \quad
  [\emph{e.g.}]\footnote{It stands for \textit{exempli gratia} in
  Latin, which means \textit{for example}.} Do 1+0 fit with 2
  parameters: \{$A_0$, $E_0$\}
\item \label{it:sb-2} Feed the previous fit results as prior
  information for the next fit: $ \lambda_p = \bar{\lambda} \,, \;
  \sigma_p(\lambda) = \sigma_p^{\max} (\lambda)$.
  We do not impose any constraint on new fit parameters
  which are not included in the previous fit.
  [\emph{e.g.}] Results for the 1+0 fit are used as prior information
  on the 1+1 fit such that $[A_0]_p = \bar{A}_0$, $\sigma_p(A_0) =
  \sigma_p^{\max}(A_0) = \sigma_p^\text{sc}(A_0)$ and $[E_0]_p =
  \bar{E}_0$, $\sigma_p(E_0) = \sigma_p^{\max}(E_0) =
  \sigma_p^\text{mf}(E_0)$ with no constraint on $R_1$ and $\Delta
  E_1$.

  In order to give a general picture of our methodology, let us
  consider the $m+n$ fit, in which we want to determine $N=2(m+n)$ fit
  parameters.
  Hence, we should select $N$ time slices such as $\{ t_1, t_2,\ldots,
  t_N \}$, to determine the initial guess $A_0^g$, $E_0^g$, $\{
  R_j^g$, $\Delta E_j^g \}$ by solving the following equations
  using the Newton method.
  \begin{align}
    q(t_j) &\equiv \frac{f(t_j) - C(t_j)}{C(t_j)} = 0 \,,
    \quad \text{with} \quad j=1,\cdots,N
    \label{eq:newton-sim-eq-N}
  \end{align}
  where $C(t)$ is data for the 2pt correlation functions in
  Eq.~\eqref{eq:corr-2pt} and $f(t)$ is the fit function in
  Eq.~\eqref{eq:fit-func}.

  \begin{enumerate}[label=\textbf{Step-2\Alph*},ref=\textbf{Step-2\Alph*},itemsep=0.05em]
  \item \label{it:a-1} We find all the possible combinations of $N$
    time slices which satisfy the following conditions.
    \begin{itemize}[itemsep=0.05em]
    \item We choose $N$ time slices within the fit range:
      $t_{\min} \le t \le t_{\max}$.
    \item $t_{\min}$ should be included.
    \item The number of even time slices should be equal to that of
      odd time slices in order to avoid a bias.
    \end{itemize}
  \item \label{it:a-2} Use the Newton method for each time
    slice combination to obtain a good initial guess for the $\chi^2$
    minimizer:
    \begin{enumerate}[label=\textbf{Step-2B\arabic*},ref=\textbf{Step-2B\arabic*},itemsep=0.05em]
    \item \label{it:n-1} Take $i$-th time slice combination ($1 \le i
      \le N_c$, $N_c$ is the total number of the time slice
      combinations).
    \item \label{it:n-2} Recycle fit results from the previous fit
      ($[m-1]+n$ fit or $m+[n-1]$ fit) to set part of the initial
      guess for the Newton method.
    \item \label{it:n-3} Use the scanning method \cite{
      Bhattacharya:2021peq} to set the remaining part of the initial
      guess for the Newton method.
    \item \label{it:n-4} Run the Newton method.
    \item \label{it:n-5} If the Newton method finds roots, then save
      them. If it fails, discard the $i$-th time slice combination.
      [\emph{e.g.}] The failure rate is about $\approx 21\%$ for the
      1+1 fit.
    \item \label{it:n-6} Take the next $\bigl($\emph{i.e.}
      ($i+1$)-th$\bigr)$ time slice combination, and repeat the loop
      (\textbf{\ref{it:n-1}}--\textbf{\ref{it:n-6}}) until we consume
      all the time slice combinations.
    \end{enumerate}
  \item \label{it:a-3} Perform the least $\chi^2$ fitting with the
    initial guess obtained by the Newton method.
  \item \label{it:a-4} Sort results for $\chi^2$, check the $\chi^2$
    distribution, and find out whether the $\chi^2$ minimizer
    converges to the global minimum or a local minimum.
  \end{enumerate}
\item \label{it:sb-3} Perform stability tests to obtain optimal prior
  widths.
  [\emph{e.g.}] Determine \{$\sigma_p^{\text{opt}}(A_0)$,
  $\sigma_p^{\text{opt}}(E_0)$\}.
\item \label{it:sb-4} Save the current fitting results ([\textit{e.g.}]
  1+1 fit) into the 1st fitting.
\item \label{it:sb-5} Take the next fitting ([\textit{e.g.}] 2+1
  fit) as the current fitting.
\item \label{it:sb-6} Go back to \textbf{\ref{it:sb-2}} and repeat the
  loop until we consume all the time slices allowed by the physical
  positivity \cite{ Luscher:1976ms, Luscher:1984is}.
  [\emph{e.g.}] 1+0 fit $\to$ 1+1 fit $\to$ 2+1 fit $\to$ 2+2 fit
  $\to$ $\cdots$.
\end{enumerate}
The stability test condition for the sequential Bayesian method is
%
\begin{enumerate}[ label=\textbf{(\roman*)},
    ref=\textbf{(\roman*)}, leftmargin=0.5cm, itemsep=0.05em ]
\item \label{it:st-1}
  We have freedom to adjust $\lambda_p$ so that
  $| \bar\lambda - \lambda_p| \ll \sigma_\sigma(\lambda)$.
  One way to achieve this criterion is $|\bar{\lambda} - \lambda_p| <
  10^{-4} \sigma(\lambda) \ll \sigma_\sigma(\lambda) $\,.
  Here the scaling factor $10^{-4}$ is obtained empirically for our
  statistical sample of $\approx 1000$ gauge configurations.

\item \label{it:st-2} $\sigma_p(\lambda)$ should not disturb
  $\sigma(\lambda)$ such that
  $| \sigma(\lambda, \sigma_p(\lambda)) - \sigma(\lambda, \max) |
  \ll \sigma_\sigma(\lambda) $\,.
\item \label{it:st-3} Based on condition \ref{it:st-1} and
  \ref{it:st-2}, we find the optimal prior width
  $\sigma^\text{opt}_p$.
  Here, $\sigma^\text{opt}_p$ is the minimum value of the prior width
  which does not disturb $\sigma(\lambda)$ of the fitting results such
  that
  \begin{align}
    \sigma_p^\text{opt} = \min(\sigma_p) \ \text{for} \
    \forall \sigma_p
    \in \{\sigma_p | \bar\lambda(\sigma_p) \circeq \lim_{\sigma_p^t \to \infty}
    \bar\lambda(\sigma_p^t), \
    \sigma(\lambda, \sigma_p) \circeq \lim_{\sigma_p^t \to \infty}
    \sigma(\lambda, \sigma_p^t) \}
    \label{eq:opt-pwidth-def}
  \end{align}
\end{enumerate}
Here the new equal symbol ($\circeq$) means that they are equal
within the statistical uncertainty of $\sigma_\sigma(\lambda)$.
Let us explain the stability test (\textbf{\ref{it:sb-3}}), using the
$2+1$ fit as an example.
In Fig.~\ref{fig:stab-2+1-1}, we describe how to set the optimal prior
widths $\sigma_p^\text{opt} (A_0)$ and $\sigma_p^\text{opt} (E_0)$ in
the $2+1$ fit.
In the case of $\sigma(A_0)$ in Fig.~\ref{fig:stab-2+1-1}
(\subref{fig:subfig-2+1-stab-A0-1}), we plot $\sigma(A_0)$ as a
function of $\sigma_p(A_0)$ in the unit of $\sigma(A_0; \max)$ while
we fix $\sigma_p(E_0)$ to its maximum value: $\sigma_p(E_0) =
\sigma_p^{\max}(E_0) = \sigma_p^\text{mf}(E_0)$.
Here we find that $\sigma_p^\text{opt}(A_0) = 20 \times [\sigma(A_0;
  \max)]$, which corresponds to the red (dashed) lines and red cross
symbol in Fig.~\ref{fig:stab-2+1-1}
(\subref{fig:subfig-2+1-stab-A0-1}).
Here the blue cross symbol and green (dashed) lines represents
$\sigma(A_0; \max)$ and the blue dotted lines represents
the error of error $\sigma_\sigma(A_0)$ of $\sigma(A_0; \max)$.

In Fig.~\ref{fig:stab-2+1-1} (\subref{fig:subfig-2+1-stab-E0-1}), we
present the same kind of a plot for $\sigma(E_0)$ with the same color
convention as Fig.~\ref{fig:stab-2+1-1}
(\subref{fig:subfig-2+1-stab-A0-1}).
Here we find that $\sigma_p^\text{opt}(E_0) = 25 \times
[\sigma(E_0;\max)]$.

%
%

%
%
\section{Application of the Newton method to the 2+2 fit}
\label{sec:app-newton}

\subsection{Example for {\ref{it:a-1}}}
\label{ssec:step-A}

As explained in Ref.~\cite{ Bhattacharya:2021peq}, we use the
multi-dimensional Newton method \cite{Press:2007nr,Broyden:1965br} to
obtain a good initial guess for the $\chi^2$ minimizer in the $m+n$
correlator fit.
For the $2+2$ fit, we should determine 8 fit parameters: $A_0$, $E_0$,
$\{ R_j$, $\Delta E_j \}$ with $j=1,2,3$.
Hence, we need an initial guess: $A_0^g$, $E_0^g$, $\{ R_j^g$, $\Delta
E_j^g \}$ ($j=1,2,3$).

In order to find an initial guess, we need 8 time slices ($T_8$) so
that we can use the Newton method to find roots:
\begin{align}
q(t_k) &\equiv  \frac{f(t_k) - C(t_k)}{C(t_k)} = 0 \,, \quad k = 1,2,\ldots,8
  \label{eq:newton-sim-eq-N-2}
\end{align}
where $t_k \in T_8$, $f(t)$ is the fitting function, and $C(t)$ is the
2pt correlator data.
The 8 time slices in $T_8$ should be chosen within the fit range
$t_{\min} \le t \le t_{\max}$ with $t_{\min} = 3$ and $t_{\max} = 30$.
It is required to set $t_1$ to $t_1 = t_{\min}$.
The number of even (odd) time slices is 14 (13) except for
$t_1=t_{\min}$.
Hence, the total number of the possible combinations for $T_8$
is 286,286.
\begin{align}
  {}_{14} C_4 \times {}_{13} C_3 &= \text{286,286}
\end{align}

\subsection{Description of {\ref{it:a-2}}}

In \textbf{\ref{it:a-2}}, we run the Newton method.
For example, in 2+2 fit, we select a time slice combination out of the
286,286 combinations [\textbf{\ref{it:n-1}}].
We need another initial guess as input to run the Newton method:
$A_0^{gn}$, $E_0^{gn}$, $\{ R_j^{gn}$, $\Delta E_j^{gn} \}$ where the
superscript ${}^{gn}$ indicates the initial guess for the Newton
method. 
We use fit results from the 2+1 fit to set up $A_0^{gn}$, $E_0^{gn}$,
$\{ R_j^{gn}$, $\Delta E_j^{gn} \}$ ($j=1,2$) [\textbf{\ref{it:n-2}}].
We use the scanning method in Ref.~\cite{Bhattacharya:2021peq} to
determine $R_3^{gn}$ and $\Delta E_3^{gn}$ [\textbf{\ref{it:n-3}}].

Now, we run the Newton method to find roots: $A_0^{g}$, $E_0^{g}$, $\{
R_j^{g}$, $\Delta E_j^{g} \}$ [\textbf{\ref{it:n-4}}].
If the Newton method finds a root, save them. Otherwise, discard it
[\textbf{\ref{it:n-5}}].
It turns out that the Newton method can find 16,574 roots out of the
286,286 combinations, while the rest fails.
We select 1,242 roots randomly out of the whole 16,574 roots in order
to monitor statistics for the $\chi^2$ distribution.
%

\subsection{Description of \ref{it:a-3} and \ref{it:a-4}}
%
For each root that the Newton method can find successfully, we use it
as an input to perform the least $\chi^2$ fitting
[\textbf{\ref{it:a-3}}].
For each root, we determine the statistics for $A_0$, $E_0$, $\{ R_j$,
$\Delta E_j \}$ ($j=1,2,3$) and $\chi^2$/d.o.f, using the jackknife
resampling.
Using the 1,242 roots, we check whether the $\chi^2$ minimizer reaches
the global minimum or local minima.

\begin{table}[!t]
  \center
  \vspace*{-5mm}
  \begin{tabular}{@{\quad}r@{\quad}|@{\quad}r@{\quad}|@{\quad}l@{\quad}|@{\quad}l@{\quad}}
    \hline
    \hline
    ID & $N_r$ & $\chi^2$/d.o.f. & note \\
    \hline
    \texttt{2+2/G}   & 1000 & 0.4091(82) & \textbf{global minimum} \\
    \texttt{2+2/L1}  & 167  & 0.6093(88) & $\Delta E_3 < 0$ \\
    \texttt{2+2/L2}  & 54    & 3.766(26) & $R_3 < O(10^{-9}) \approx 0$ or $R_3 < 0$ \\
    \hline
    \hline
  \end{tabular}
  \caption{Patterns of the $\chi^2$ distribution. Here, $N_r$ means
    the number of roots obtained by the Newton method.  Here
    \texttt{G} represents the global minimum and \texttt{L} represents
    the local minima.  }
  \label{tab:2+2-minima}
\end{table}

We summarize patterns for the $\chi^2$ distribution in the Table
\ref{tab:2+2-minima}.
Among 1,242 roots, 1,000 roots converges to the global minimum
(pattern ID = \texttt{2+2/G}).
We find two local minima of $\chi^2$: the pattern ID = \texttt{2+2/L1}
(167 roots) and the pattern ID = \texttt{2+2/L2} (54 roots).
The \texttt{2+2/L1} pattern gives consistent results of $\Delta E_3
= -0.409(55) < 0$, which is definitely unphysical and wrong.
The \texttt{2+2/L2} pattern gives consistent results of $R_3 <
O(10^{-9}) \approx 0$ or $R_3 < 0$, which are unphysical and wrong.
There are 21 values of the $\chi^2$/d.o.f. between the \texttt{2+2/L1}
and \texttt{2+2/L2} patterns, which also gives wrong results for $R_3$
or $\Delta E_3$.
Table \ref{tab:2+2-minima} shows that we can find the global minimum
of the $\chi^2$ distribution with the Newton method reliably.
In addition, we find that the local minima of the $\chi^2$
distribution always come up with unphysical (= wrong) results for
$R_3$ or $\Delta E_3$.

\begin{figure}[t!]
  \centering
  \begin{subfigure}{0.49\textwidth}
    \vspace*{-5mm}
    \includegraphics[width=\linewidth]{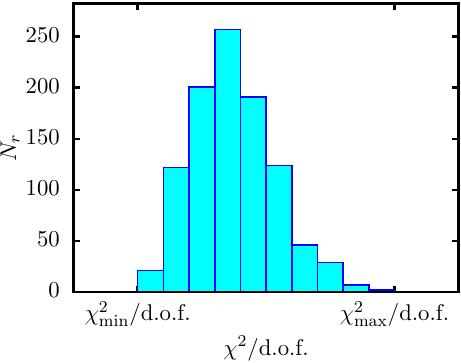}
    \caption{Histogram of the $\chi^2_d$ distribution.  We use the
      same notation as in Table \ref{tab:2+2-minima}. Here $\chi^2_d
      \equiv \chi^2/\text{d.o.f.}$, $\chi^2_{\min} \equiv
      \min(\chi^2)$, and $\chi^2_{\max} \equiv \max(\chi^2)$.}
    \label{fig:subfig-2+2-Bm-histogram}
  \end{subfigure}
  \hfill
  \begin{subfigure}{0.49\textwidth}
    \vspace*{-5mm}
    \includegraphics[width=\linewidth]{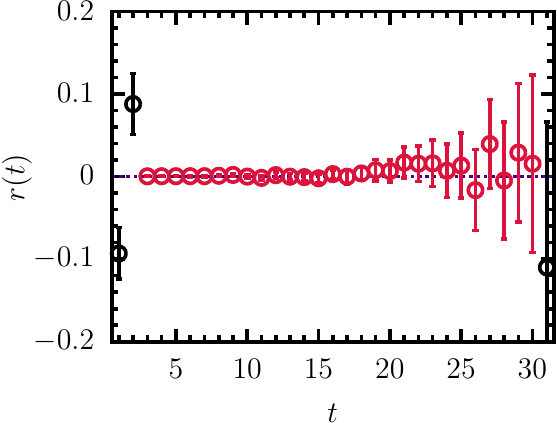}
    \caption{Residual $r(t)=\dfrac{f(t)-C(t)}{|C(t)|}$.  Here the red
      color represents the fit range ($3 \le t \le 30$).  The errors
      are purely statistical.}
    \label{fig:subfig-2+2-Bm-resid}
  \end{subfigure}
  \caption{$\chi^2$ distribution and residual coming from the data
    analysis on the $B$-meson propagators.}
  \label{fig:2+2-hst-resid}
\end{figure}

In Fig.~\ref{fig:2+2-hst-resid}
(\subref{fig:subfig-2+2-Bm-histogram}), we show the histogram of the
$\chi^2$/d.o.f.~distribution of the \texttt{2+2/G} pattern.
This indicates that the $\chi^2$ minimizer find the global minimum
with about $\approx 80\%$ probability which indicates that once out of
five times the $\chi^2$ minimizer converges to local minima.
Hence, this is a clear advantage of using the Newton method to find
a good initial guess for the $\chi^2$ minimizer.
Out of the set of multiple roots of the Newton method, a subset find
the global minimum for the $\chi^2$ distribution, and another subset
reach the local minima, which we can discard without loss of
generality.

In Fig.~\ref{fig:2+2-hst-resid}
(\subref{fig:subfig-2+2-Bm-histogram}), $\Delta \chi^2/\text{d.o.f.}
= \chi^2_{\max}/\text{d.o.f.} - \chi^2_{\min}/\text{d.o.f.} \cong
10^{-8}$ is significantly less than the statistical error ($\cong
0.0082$) of the $\chi^2 / \text{d.o.f.}$.
This indicates that the $\chi^2$ distribution has sharp peak.

The residual plot of $B$ meson 2pt correlator for the $2+2$ fit is
given in Fig.~\ref{fig:2+2-hst-resid}
(\subref{fig:subfig-2+2-Bm-resid}).
Note that we consumed up all the possible time slices allowed by
the physical positivity \cite{ Luscher:1976ms, Luscher:1984is}.


%
%
\section{Preliminary result on $h_{A_1}(w=1)/\rho_{A_1}$: $B \to D^\ast \ell \nu$ form factor at zero recoil}
\label{sec:hA1}

The semileptonic form factor $h_{A_1}(w)$ at zero recoil (\emph{i.e.}
$w=1$) can be obtained with the Hashimoto ratio \cite{Hashimoto:1999yp}:
\begin{align}
  h_{A_1}(w=1) & = \rho_{A_1} \sqrt{ \frac{\langle D^\ast_0 | A^{cb}_j
      | {B}_0 \rangle_{\text{L}} \, \langle {B}_0 | A^{bc}_j | D^\ast_0
      \rangle_{\text{L}}}{\langle D^\ast_0 | V^{cc}_4 | D^\ast_0 \rangle_{\text{L}} \,
      \langle {B}_0 | V^{bb}_4| {B}_0 \rangle_{\text{L}}} } \,,
  \qquad \rho_{A_1} = \sqrt{\frac{Z_A^{cb} \, Z_A^{bc}}{Z_V^{cc} \,
      Z_V^{bb}}} \,.
  \label{eq:h_A1-def}
\end{align}
The one-loop matching calculation of $\rho_{A_1}$ is underway
\cite{Bailey:2023sk}.
Here we present preliminary results on blind
(\emph{i.e.}~$\rho_{A_1}=1$) $h_{A_1}(w = 1)$ in this work.
The subscript ${}_{0}$ in Eq.~\eqref{eq:h_A1-def} represents the
ground states at zero momentum ($\vec{p}_X = 0$ with $X =
B,\ D^\ast$).
We want to extract the four ground state matrix elements: $\langle
D^\ast_0 | A^{cb}_j | {B}_0 \rangle_{\text{L}}$, $\langle {B}_0 |
A^{bc}_j | D^\ast_0 \rangle_{\text{L}}$, $\langle D^\ast_0 | V^{cc}_4
| D^\ast_0 \rangle_{\text{L}}$, and $\langle {B}_0 | V^{bb}_4| {B}_0
\rangle_{\text{L}}$ from the 3pt correlation functions.
The 3pt correlation functions are calculated on the lattice and so the
Hilbert space consists in quark and gluon states.
The Hilbert space for physical observables such as the matrix elements
consists in hadronic states.
For example, when we fit the 2pt correlation functions for $B$-meson
propagators, we obtain results for $A_0$, $E_0$
(\emph{i.e.}~information on the $B_0$ ground state), $R_1$, $\Delta
E_1$ (\emph{i.e.}~the $B_1$ excited state with odd time-parity),
$R_2$, $\Delta E_2$ (\emph{i.e.}~the $B_2$ excited state with even
time-parity) and so on.
We can obtain similar results for $D^\ast$-meson propagators.
The fitting functional form for the 3pt correlation functions
calculated on the lattice in the $B \to D^\ast$ channel is
\begin{align}
  f_{T_{\text{sep}}}^{B\to D^{\ast}}\left(t\right)
  & = \left\langle D_{0}^{\ast}\right| A_{j}^{cb}
  \left|B_{0}\right\rangle_{\text{L}} \, k_{0}^{D^{\ast}}\left(t\right)
  k_{0}^{B}\left(T_{\text{sep}}-t\right)
  + \left\langle D_{0}^{\ast}\right| A_{j}^{cb}
  \left|B_{2}\right\rangle_{\text{L}} \, k_{0}^{D^{\ast}}\left(t\right)
  k_{2}^{B}\left(T_{\text{sep}}-t\right) \nonumber \\
  & \hphantom{=\ } + \left\langle D_{1}^{\ast}\right| A_{j}^{cb}
  \left|B_{1}\right\rangle_{\text{L}} \, k_{1}^{D^{\ast}}\left(t\right)
  k_{1}^{B}\left(T_{\text{sep}}-t\right)
  + \left\langle D_{2}^{\ast}\right| A_{j}^{cb}
  \left|B_{0}\right\rangle_{\text{L}} \, k_{2}^{D^{\ast}}\left(t\right)
  k_{0}^{B}\left(T_{\text{sep}}-t\right) \nonumber \\
  & \hphantom{=\ } + \left\langle D_{1}^{\ast}\right| A_{j}^{cb}
  \left|B_{3}\right\rangle_{\text{L}} \, k_{1}^{D^{\ast}}\left(t\right)
  k_{3}^{B}\left(T_{\text{sep}}-t\right)
  + \left\langle D_{2}^{\ast}\right| A_{j}^{cb}
  \left|B_{2}\right\rangle_{\text{L}} \, k_{2}^{D^{\ast}}\left(t\right)
  k_{2}^{B}\left(T_{\text{sep}}-t\right) \nonumber \\
  & \hphantom{=\ } + \left\langle D_{3}^{\ast}\right| A_{j}^{cb}
  \left|B_{1}\right\rangle_{\text{L}} \, k_{3}^{D^{\ast}}\left(t\right)
  k_{1}^{B}\left(T_{\text{sep}}-t\right)
  + \left\langle D_{3}^{\ast}\right| A_{j}^{cb}
  \left|B_{3}\right\rangle_{\text{L}} \, k_{3}^{D^{\ast}}\left(t\right)
  k_{3}^{B}\left(T_{\text{sep}}-t\right) \,, \\
  k_{0}^{X}\left(t\right) & = \sqrt{A_{0}^{X}}\,e^{-E_{0}^{X}\,t} \,,
  \qquad \qquad \quad
  k_{1}^{X}\left(t\right) = \sqrt{A_{0}^{X}
    R_{1}^{X}}\,e^{-\left(E_{0}^{X}+\Delta
    E_{1}^{X}\right)\,t}\left(-1\right)^{t+1} \,, 
  \nonumber \\
  k_{2}^{X}\left(t\right) & = \sqrt{A_{0}^{X}
    R_{2}^{X}}\,e^{-\left(E_{0}^{X}+\Delta E_{2}^{X}\right)\,t} \,,
  \quad
  k_{3}^{X}\left(t\right) = \sqrt{A_{0}^{X} R_{1}^{X}
    R_{3}^{X}}\,e^{-\left(E_{0}^{X}+\Delta E_{1}^{X}+\Delta
    E_{3}^{X}\right)\,t}\left(-1\right)^{t+1} \,.
  \label{eq:3pt-ff}
\end{align}
The $k_j^X(t)$ ($j=0,1,2,3$, $X=B,D^\ast$) comes from the fit results
for the 2pt correlation functions for the $B$ and $D^\ast$ mesons.
Hence, we determine the lattice matrix elements simply by a linear
fit.
As a result, we obtain $\langle D^\ast_0 | A^{cb}_j | {B}_0
\rangle_{\text{L}}$.
We can apply the same kind of fitting to the $D^\ast \to B$, $B \to
B$, and $D^\ast \to D^\ast$ channels.
As a results, we obtain the rest of the lattice matrix elements:
$\langle {B}_0 | A^{bc}_j | D^\ast_0 \rangle_{\text{L}}$, $\langle
D^\ast_0 | V^{cc}_4 | D^\ast_0 \rangle_{\text{L}}$, $\langle {B}_0 |
V^{bb}_4| {B}_0 \rangle_{\text{L}}$.

\begin{figure}[t!]
  \centering
  \vspace*{-8mm}
  \includegraphics[width=0.7\linewidth]{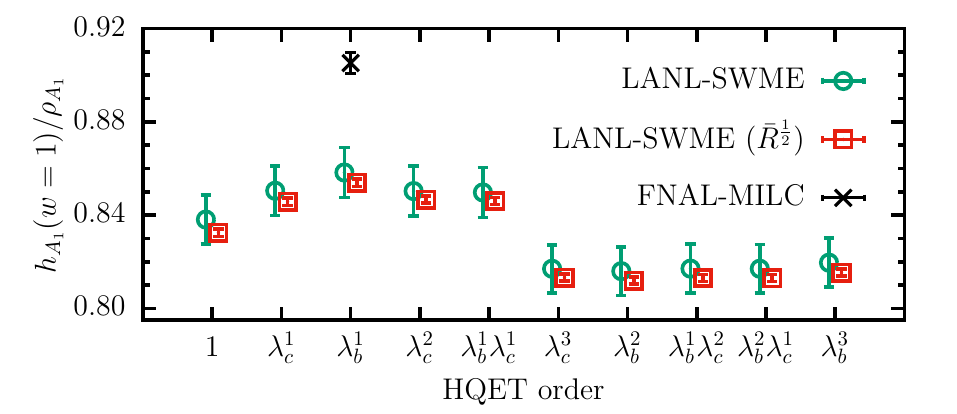}
  \caption{ $h_{A_1}(w=1)/\rho_{A_1}$ in the order of the HQET power
    counting. Here $\lambda_c \cong 1/5$, and $\lambda_b \cong 1/17$. }
  \label{fig:hA1-result}
\end{figure}

In Fig.~\ref{fig:hA1-result} we present results for
$h_{A_1}(w=1)/\rho_{A_1}$.
Here the green circles represent our results for
$h_{A_1}(w=1)/\rho_{A_1}$ with no contamination from exited states,
while the red squares represent those obtained using the $\bar{R}$
ratio \cite{FermilabLattice:2014ysv} which include some contamination
from excited states by construction.
The black cross represents the FNAL-MILC result for
$h_{A_1}(w=1)/\rho_{A_1}$ which is obtained using the $\bar{R}$ ratio
with the Fermilab action for bottom and charm quarks, and the asqtad
action for light quarks with $N_f = 2+1$
\cite{FermilabLattice:2014ysv}.

When we do the linear fit over the 3pt correlation functions, we could
not use the full covariance fitting but the diagonal approximation
\cite{Yoon:2011wdw} due to unwanted bias by strong correlation between
different time slices.
We find that off-diagonal elements of the correlation matrix
$\rho(t_i,t_j)$ with $t_i \ne t_j$ are close to one.
This issue needs further investigation.
%
%

\acknowledgments

We would like to thank Andreas Kronfeld, and Carlton Detar for helpful
discussion on many issues on theory and fitting.
We would like to thank the MILC collaboration and Chulwoo Jung for
providing the HISQ lattice ensembles to us.
The research of W.~Lee is supported by the Mid-Career Research Program
(Grant No. NRF-2019R1A2C2085685) of the NRF grant funded by the Korean
government (MSIT).
W.~Lee would like to acknowledge the support from the KISTI
supercomputing center through the strategic support program for the
supercomputing application research (No. KSC-2018-CHA-0043,
KSC-2020-CHA-0001, KSC-2023-CHA-0010).
Computations were carried out in part on the DAVID cluster at Seoul
National University.

\bibliography{ref}

\end{document}